\documentclass[prc,twocolumn,superscriptaddress,showpacs,preprintnumbers,amsmath,amssymb]{revtex4-1}
\usepackage[dvips]{graphicx}
\usepackage{dcolumn}
\usepackage{bm}

\begin{document}
\title{Number conservation in odd-particle number random phase approximation and extensions}
\author{Mitsuru Tohyama}
\affiliation{Faculty of Medicine, Kyorin University, Mitaka, Tokyo
181-8611, Japan}

%\date{\today} 
\begin{abstract}
The number conservation law in
the odd-particle number random-phase approximation (oRPA) and its extension (EoRPA) is studied by applying them to a pairing model and $^{16}$O. 
It is found in the application to $^{16}$O that the number conservation law is not fulfilled in oRPA and EoRPA and that it is drastically improved in EoRPA due to 
the inclusion of ground-state correlation effects.
\end{abstract}
\maketitle
\section{Introduction}
The standard approach to study the single-particle properties of atomic nuclei may be the Green's function method \cite{RS,s21}.
Various theoretical approaches have been proposed to implement the coupling to low-lying collective modes into the self-energy of the Green's function: For example, the particle-vibration coupling (PVC) models  \cite{Giai,bortignon,litv,colo},
the Tamm-Dancoff approximation (TDA) \cite{hrpa96} and the Faddeev random-phase approximation (FRPA) \cite{dick02}.
The odd-particle number random-phase approximation (oRPA) and its extension (EoRPA) have also been used to study the single-particle properties of $^{16}$O \cite{ts13}.
oRPA and EoRPA provide us alternative approaches to calculate the single-particle Green's function: The single-particle Green's function is expressed by using the eigenvalues and 
the transition amplitudes in oRPA and EoRPA. In the Green's function method the occupation probabilities of the single-particle states are obtained from the residues of its poles \cite{RS}.
The conservation law of the total number of particles requires that the number of particles depleted from the single-particle states 
below the Fermi level be the same as the number of particles occupying the single-particle states above the Fermi level. This is also known as the Luttinger theorem \cite{lutt2}
which has been attracting much theoretical and experimental interests in the fields of condensed-matter physics \cite{heath} and atomic physics \cite{sogo,urban,pieri,durel}.
The proof of this theorem is given in Refs. \cite{lutt2,heath} for the exact theory and 
a question is what happens to it when approximations are made in realistic calculations
where diagrams are partially summed up in any case. This problem has been addressed in TDA \cite{hrpa96} and FRPA \cite{dick02} for finite nuclei. 
It has also been  discussed for nuclear matter cases \cite{ramos,mahaux}. 
In this paper the number conservation law is studied for oRPA and EoRPA. EoRPA is formulated based a correlated ground state 
given by the time-dependent density-matrix theory (TDDM) \cite{toh20}.
It is pointed out that the neglect of the coupling to higher amplitudes in oRPA and EoRPA inevitably brings about the violation
of the Luttinger theorem
and it is demonstrated that the effects of ground-state correlations included in EoRPA drastically improves the number conservation law.  
The paper is organized as follows: The formulation of EoRPA is given in Sect. II, the properties of EoRPA and its relations to other approaches are discussed in Sect. III
and the applications of oRPA and EoRPA to a pairing model \cite{rich} and $^{16}$O are presented in Sect.IV. Section V is devoted to summary.

\section{Formulation}
Let us consider a nucleus consisting of $N$ nucleons and assume that the 
total Hamiltonian $H$
consists of the kinetic energy term and a two-body interaction. Let us assume 
that $|0\rangle$ is the ground state of the $N$ nucleon system with  even $N$ 
and energy $E_0$ and that 
$|\mu\rangle$ is an exact eigenstate of the Hamiltonian 
for the $N-1$ system with an eigenvalue $E_\mu$ ($H|\mu\rangle=E_\mu|\mu\rangle$).
(As will be pointed out below, oRPA and EoRPA have particle-hole symmetry and can describe the $N+1$ system as well.)

\subsection{Single-paticle Green's function and transition amplitudes}
First it is explained that the single-particle Green's function is expressed with the transition amplitudes $x^\mu_\alpha=\langle 0|a^+_{\alpha}|\mu\rangle$
and $\chi^\mu_\alpha=\langle \mu|a_{\alpha}|0\rangle$, where $a^+_{\alpha}~(a_{\alpha})$ is the creation (annihilation) operator of a nucleon at a single-particle state $\alpha$.
In the exact case $x^\mu_\alpha=(\chi^\mu_\alpha)^*$ holds.
The Green's function $G_{\alpha\alpha'}(t-t')$ for time $t>t'$ is given by
\begin{eqnarray}
G_{\alpha\alpha'}(t-t')=i\langle 0|a^+_{\alpha'}(t)a_\alpha(t')|0\rangle,
\end{eqnarray}
where $a^+_{\alpha}(t)$ is $e^{iHt}a^+_\alpha e^{-iHt}$. Units such as $\hbar=1$ are used hereafter. The Green's function is written with $x^\mu_\alpha$ and $\chi^\mu_\alpha$ as 
\begin{eqnarray}
G_{\alpha\alpha'}(t-t')&=&i\langle 0|a^+_{\alpha'}(t)a_\alpha(t')|0\rangle
\nonumber \\
&=&i\sum_\mu\langle 0|e^{iHt}a^+_{\alpha'} e^{-iHt}|\mu\rangle\langle\mu|e^{iHt'}a_\alpha e^{-iHt'}|0\rangle
\nonumber \\
&=&i\sum_\mu x^\mu_{\alpha'}\chi^\mu_\alpha e^{i\omega_\mu (t-t')},
\end{eqnarray}
where $\omega_\mu=E_0-E_\mu$. The occupation matrix $n_{\alpha\alpha'}$ is given by \cite{RS}
\begin{eqnarray}
n_{\alpha\alpha'}=-i\lim_{t\rightarrow t'}G_{\alpha\alpha'}(t-t')
=\sum_{\mu}x^\mu_{\alpha'}\chi^\mu_\alpha,
\end{eqnarray}
and in the exact case $\sum_\alpha n_{\alpha\alpha}$ gives $N$.

\subsection{Time-dependent density-matrix theory and number conservation}
As mentioned above, EoRPA is formulated based on the TDDM ground state. In the following it is shown that TDDM conserves the total number of particles. 
TDDM consists of the coupled equations of motion for 
the occupation matrix $n_{\alpha\alpha'}$ and the two-body correlation matrix $C_{\alpha\beta\alpha'\beta'} (C_2)$ defined by  
\begin{eqnarray}
n_{\alpha\alpha'}&=&\langle 0|a^+_{\alpha'}a_\alpha|0\rangle,
\label{n}
\\
C_{\alpha\beta\alpha'\beta'}&=&\langle 0|:a^+_{\alpha'}a^+_{\beta'}a_\beta a_\alpha:|0\rangle,
\label{C2}
\end{eqnarray}
where  $:~:$ implies that lower-order operators are subtracted such that
\begin{eqnarray}
:a^+_{\alpha'}a^+_{\beta'}a_\beta a_\alpha:&=&a^+_{\alpha'}a^+_{\beta'}a_\beta a_\alpha-{\cal A}(n_{\alpha\alpha'}n_{\beta\beta'})
\nonumber \\
&-&{\cal AS}(n_{\alpha\alpha'}a^+_{\beta'}a_\beta).
\end{eqnarray}
Here, ${\cal A}()$ (${\cal S}()$)  antisymmetrizes (symmetrizes) the quantities in parentheses under the exchange of the single-particle indices.
The TDDM equations have been used to calculate the correlated ground states of closed-shell nuclei using an adiabatic method \cite{toh20}.
The TDDM equations conserve the total number of particles in the adiabatic method,  
where the residual interaction $V$ gradually increases as a function of time $t$ such that
$H(t)=H_0+t/T\times V$. Here it is assumed that $T$ is large and that the correlated ground state wavefunction is given by $|\Psi_0(t)\rangle= \exp[-i\int_0^t Hdt]|\Psi_0\rangle$, where $|\Psi_0\rangle$
is the ground state of $H_0$ (the Hartree-Fock (HF) ground state is usually used for $|\Psi_0\rangle$). This adiabatic method is based on the Gell-mann-Low theorem \cite{gell}.
The equation of motion for $n_{\alpha\alpha'}$ is obtained by replacing $|0\rangle$ in Eqs. (\ref{n}) and (\ref{C2}) with $|\Psi_0(t)\rangle$
\begin{eqnarray}
i\frac{dn_{\alpha\alpha'}}{dt}&=&\langle \Psi_0(t)|[a^+_{\alpha'}a_\alpha,H]| \Psi_0(t)\rangle
\nonumber \\
&=&(\epsilon_\alpha(t)-\epsilon_{\alpha'}(t))n_{\alpha\alpha'}(t)
\nonumber \\
&+&\sum_{\lambda_1\lambda_2\lambda_3}[\langle\alpha\lambda_3|v(t)|\lambda_1\lambda_2\rangle C_{\lambda_1\lambda_2\alpha'\lambda_3}(t)
\nonumber \\
&-& C_{\alpha\lambda_3\lambda_1\lambda_2}(t)\langle\lambda_1\lambda_2|v(t)|\alpha'\lambda_3\rangle],
\label{lambda1}
\end{eqnarray}
where $v(t)=t/T\times V$ and the single-particle energy matrix
\begin{eqnarray}
\epsilon_{\alpha\alpha'}=\epsilon^0_\alpha\delta_{\alpha\alpha'}+\sum_{\lambda\lambda'}\langle\alpha\lambda|v(t)|\alpha'\lambda'\rangle_An_{\lambda'\lambda}
\label{spenergy}
\end{eqnarray}
is assumed diagonal to simplify the equation:
In the case of a spherical nucleus the off-diagonal parts of $\epsilon_{\alpha\alpha'}$ are non-vanishing only for the single-particle states 
with different principal quantum numbers and are usually much smaller than the diagonal parts.
In Eq. (\ref{spenergy}) $\epsilon^0_\alpha$ is the HF single-particle energy.
The equation of motion for $C_2$ couples to the correlated part ($C_3$) of a three-body density matrix.
To obtain a closed set of the equations for $n_{\alpha\alpha'}$ and $C_2$, $C_3$ is usually approximated by the squares of $C_2$ \cite{toh20}.
As easily seen from
Eq. (\ref{lambda1}), the total number of particles $N=\sum_\alpha n_{\alpha\alpha}(t)$ is conserved, that is, $dN/dt=\sum_\alpha dn_{\alpha\alpha}/dt=0$
because the second term on the right-hand side of Eq. (\ref{lambda1}) vanishes when the sum is taken over all single-particle indices.
Thus TDDM fulfills the number conservation low regardless of the approximations for $C_3$.

\subsection{Equations for transition amplitudes and total number conservation}
In oRPA and EoRPA the single-fermion transition amplitudes are used to calculated the occupation probabilities of the single-particle states. 
By using the above time-dependent Hamiltonian $H(t)$ the conditions that the transition amplitudes should satisfy to fulfill the 
number conservation law are discussed below.
For this purpose the time-dependent transition amplitudes for $H(t)$ are considered
\begin{eqnarray}
x^\mu_\alpha(t)&=&\langle \Psi_0(t)|a^+_\alpha|\Psi_\mu(t)\rangle, \\
X^\mu_{\alpha\beta:\gamma}(t)&=&\langle \Psi_0(t)|:a^+_\alpha a^+_\beta a_\gamma:|\Psi_\mu(t)\rangle,
\end{eqnarray}
and
\begin{eqnarray}
\chi^\mu_\alpha(t)&=&\langle \Psi_\mu(t)|a_\alpha|\Psi_0(t)\rangle, \\
{\cal X}^\mu_{\alpha\beta:\gamma}(t)&=&\langle \Psi_\mu(t)|:a^+_\gamma a_\beta a_\alpha:|\Psi_0(t)\rangle,
\end{eqnarray}
where $|\Psi_\mu(t)\rangle= \exp[-i\int_0^t Hdt]|\Psi_\mu(0)\rangle$. Here, $|\Psi_\mu(0)\rangle$ is the eigenstate of $H_0$ with the number $N-1$.
The time derivatives of $x^\mu_\alpha(t)$ and $\chi^\mu_\alpha(t)$  give
\begin{eqnarray}
i\frac{dx^\mu_\alpha}{dt}&=&\langle \Psi_0(t)|[a^+_{\alpha},H]| \Psi_\mu(t)\rangle
\nonumber \\
&=&-\epsilon_\alpha (t)x^\mu_\alpha(t)
\nonumber \\
&+&\sum_{\lambda_1\lambda_2\lambda_3}\langle\lambda_1\lambda_2|v(t)|\alpha\lambda_3\rangle
X^\mu_{\lambda_1\lambda_2:\lambda_3}(t),
\\
i\frac{d\chi^\mu_\alpha}{dt}&=&\langle \Psi_\mu(t)|[a_{\alpha},H]| \Psi_0(t)\rangle
\nonumber \\
&=&\epsilon_\alpha(t) \chi^\mu_\alpha (t)
\nonumber \\
&-&\sum_{\lambda_1\lambda_2\lambda_3}\langle\alpha\lambda_3|v(t)|\lambda_1\lambda_2\rangle
{\cal X}^\mu_{\lambda_1\lambda_2:\lambda_3}(t).
\label{chi}
\end{eqnarray}
The occupation matrix is defined by 
\begin{eqnarray}
n_{\alpha\alpha'}(t)&=&\langle\Psi_0(t)|a^+_{\alpha'}a_\alpha|\Psi_0(t)\rangle
\nonumber \\
&=&\sum_\mu\langle\Psi_0(t)|a^+_{\alpha'}|\Psi_\mu(t)\rangle\langle\Psi_\mu(t)|a_\alpha|\Psi_0(t)\rangle
\nonumber \\
&=&\sum_\mu x^\mu_{\alpha'}(t)\chi^\mu_\alpha(t).
\end{eqnarray}
The time derivative of the total number $N=\sum_\alpha n_{\alpha\alpha}$ is given by
\begin{eqnarray}
i\frac{dN}{dt}&=i&\sum_\alpha\left(\frac{dx^\mu_\alpha}{dt}\chi^\mu_\alpha+x^\mu_\alpha\frac{d\chi^\mu_\alpha}{dt}\right)
\nonumber \\
&=&\sum_{\mu\alpha\lambda_1\lambda_2\lambda_3}[\langle\lambda_1\lambda_2|v|\alpha\lambda_3\rangle
X^\mu_{\lambda_1\lambda_2:\lambda_3}\chi^\mu_\alpha
\nonumber \\
&-&\langle\alpha\lambda_3|v|\lambda_1\lambda_2\rangle
x^\mu_\alpha{\cal X}^\mu_{\lambda_1\lambda_2:\lambda_3}],
\label{Ndot}
\end{eqnarray}
where the notation $(t)$ is omitted for simplicity.
From Eq. (\ref{Ndot}) it is found that the necessary condition for number conservation is 
\begin{eqnarray}
\sum_\mu X^\mu_{\alpha'\beta':\beta}\chi^\mu_\alpha=\sum_\mu x^\mu_{\alpha'}{\cal X}^\mu_{\alpha\beta:\beta'}.
\label{cond0}
\end{eqnarray}
The quantities on the both sides of the above equation correspond to $C_{\alpha\beta\alpha'\beta'}$ in Eq. (\ref{lambda1}). 
Equation (\ref{cond0}) requires that $C_{\alpha\beta\alpha'\beta'}$ consisting of the one-fermion and three-fermion transition amplitudes should be independent of how it is composed.
When the conditions
\begin{eqnarray} 
\chi_\alpha^\mu=(x_\alpha^\mu)^*
\label{x}
\end{eqnarray}
and
\begin{eqnarray}
{\cal X}_{\alpha\beta:\gamma}^\mu=(X_{\alpha\beta:\gamma}^\mu)^*
\label{X}
\end{eqnarray}
hold, Eq. (\ref{cond0}) is expressed with $x^\mu_{\alpha}$ and $X^\mu_{\alpha'\beta':\beta}$ as 
\begin{eqnarray}
\sum_\mu X^\mu_{\alpha'\beta':\beta}(x^{\mu }_\alpha)^*=\sum_\mu x^\mu_{\alpha'}({X}^{\mu }_{\alpha\beta:\beta'})^*.
\label{cond1}
\end{eqnarray}
This is the relation that is satisfied by exact solutions.
As will be discussed below, Eqs. (\ref{x}) and (\ref{X}) are satisfied both in oRPA and in EoRPA.
It will be also pointed out that the neglect of higher amplitudes in the 
calculations of $X_{\alpha\beta:\gamma}^\mu$ causes the violation of Eq. (\ref{cond1}).

\subsection{Equations for transition amplitudes I}
Now the equations for the transition amplitudes $x^\mu_\alpha$ and $X^\mu_{\alpha\beta:\gamma}$ that have been used in realistic calculations \cite{ts13} are presented.
Here the time-independent Hamiltonian $H$ consisting of the HF part $H_0$ and the residual interaction $v$ is considered.
The transition amplitudes are defined by
\begin{eqnarray}
x^\mu_\alpha&=&\langle 0|a^+_\alpha|\mu\rangle, \\
X^\mu_{\alpha\beta:\gamma}&=&\langle 0|:a^+_\alpha a^+_\beta a_\gamma:|\mu\rangle,
\end{eqnarray}
where $|0\rangle$ and $|\mu\rangle$ are the ground and excited states of $H$, respectively.
The occupation matrix $n_{\alpha\alpha'}$ is given by
\begin{eqnarray} 
n_{\alpha\alpha'}=\langle 0|a^+_{\alpha'}a_\alpha|0\rangle.
\end{eqnarray}
From the equation-of-motion (EoM) relation \cite{ts13}
\begin{eqnarray}
\langle 0|[H,a^+_\alpha]=\langle 0|a^+_\alpha(E_0-H),
\end{eqnarray}
the equation for $x^\mu_\alpha$ is obtained as
\begin{eqnarray}
\langle 0|[H,a^+_\alpha]|\mu\rangle&=&\omega_\mu\langle 0|a^+_\alpha|\mu \rangle=\omega_\mu x^\mu_\alpha, 
\label{eq1}
\end{eqnarray}
where $\omega_\mu=E_0-E_\mu$. Here, $E_0$ and $E_\mu$ are respectively the energies of $|0\rangle$ and $\mu\rangle$.
The commutator on the left-hand side of the above equation includes terms with $a^+_\alpha a^+_\beta a_\gamma$.
Therefore, $x^\mu_\alpha$ couples to $X^\mu_{\alpha\beta:\gamma}$. In a way 
analogous to that used in deriving Eq.~(\ref{eq1}) 
the equation for $X^\mu_{\alpha\beta:\gamma}$ is given as
\begin{eqnarray}
\langle 0|[H,:a^+_\alpha a^+_\beta a_\gamma:]|\mu \rangle&=&\omega_\mu\langle 0|:a^+_\alpha a^+_\beta a_\gamma:|\mu\rangle
\nonumber \\
&=&\omega_\mu X^\mu_{\alpha\beta:\gamma}.
\label{eq2}
\end{eqnarray}
When the coupling to higher amplitudes is neglected to close the equations within $x^\mu_\alpha$ and $X^\mu_{\alpha\beta:\gamma}$,
Eqs. (\ref{eq1}) and (\ref{eq2}) are written in matrix form \cite{ts13}:
\begin{eqnarray}
\left(
\begin{array}{cc}
a&c\\
b&d
\end{array}
\right)\left(
\begin{array}{c}
x^\mu\\
X^\mu
\end{array}
\right)
=\omega_\mu
\left(
\begin{array}{c}
x^\mu\\
X^\mu
\end{array}
\right),
\label{hrpa}
\end{eqnarray}
where the single-particle indices in $x^\mu_\alpha$ and $X^\mu_{\alpha\beta:\gamma}$ are omitted for simplicity.
The matrices $a$, $b$, $c$ and $d$ are given in Appendix A. The effects of ground-state correlations are included in $b$ and $d$ through
$n_\alpha$ and $C_2$.
For later use
the equation for $x^\mu_\alpha$ is  explicitly given 
\begin{eqnarray}
(\epsilon_\alpha -\omega_\mu)x^\mu_\alpha&+&\sum_{\lambda_1\lambda_2\lambda_3}\langle\lambda_1\lambda_2|v|\alpha\lambda_3\rangle
X^\mu_{\lambda_1\lambda_2:\lambda_3}=0.
\label{stddm1}
\end{eqnarray}

In a similar way the equations for the transition amplitudes
\begin{eqnarray}
\chi^\mu_\alpha&=&\langle \mu|a_\alpha|0\rangle, \\
{\cal X}^\mu_{\alpha\beta:\gamma}&=&\langle \mu|:a^+_\gamma a_\beta a_\alpha:|0\rangle
\end{eqnarray}
can be derived.
Since the Hamiltonian matrix of Eq. (\ref{hrpa}) is not hermitian, the eigenvalues of Eq. (\ref{hrpa}) can be complex. For these complex solutions the relation
$(\chi^\mu,{\cal X}^\mu)=((x^\mu)^*,(X^\mu)^*)$ does not hold.

\subsection{Equations for transition amplitudes II: Equation of motion approach with an excitation operator}
Equation (\ref{hrpa}) lacks some effects of higher configurations such as self-energy contributions in the configuration $X^\mu_{\alpha\beta:\gamma}$,
which should be included when a correlated ground state is used. 
Another formulation which is based on EoM \cite{Rowe,s21} is presented to take account of such effects.
By introducing the excitation operator $q^+_\mu$  
\begin{eqnarray}
q^+_\mu=\sum_\alpha y_\alpha^\mu~ a_\alpha 
+\sum_{\alpha\beta\gamma}Y_{\alpha\beta:\gamma}^\mu:a^+_\gamma a_\beta a_\alpha:
\label{operh}
\end{eqnarray}
and assuming, as usual, $q^+_\mu|0\rangle=|\mu\rangle$ and $q_\mu|0\rangle=0$,
the equations for $y^\mu$ and $Y^\mu$ are obtained from Eqs. (\ref{eq1}) and (\ref{eq2}). They are written in matrix form as 
\begin{eqnarray}
\left(
\begin{array}{cc}
A&C\\
B&D
\end{array}
\right) 
\left(
\begin{array}{c}
y^{\mu}\\
Y^{\mu}
\end{array}
\right)
=\omega_\mu
\left(
\begin{array}{cc}
N_{11}&N_{12}\\
N_{21}&N_{22}
\end{array}
\right)
\left(
\begin{array}{c}
y^{\mu}\\
Y^{\mu}
\end{array}
\right),
\label{hrpa1}
\end{eqnarray}
where the matrices are defined as
\begin{eqnarray}
A(\alpha:\alpha')&=&\langle 0|\{[H,a^+_{\alpha}],a_{\alpha'}\}|0\rangle, \\
B(\alpha\beta\gamma:\alpha') &=&\langle 0|\{[H,:a^+_\alpha a^+_\beta a_\gamma:], a_{\alpha'}\}|0\rangle,\\
C(\alpha:\alpha'\beta'\gamma')&=&\langle 0|\{[H,a^+_{\alpha}:],:a^+_{\gamma'} a_{\beta'}a_{\alpha'} \}|0\rangle,
\\
D(\alpha\beta\gamma:\alpha'\beta'\gamma'&)&
\nonumber \\
=\langle 0|\{[H&,&:a^+_\alpha a^+_\beta a_\gamma:],:a^+_{\gamma'} a_{\beta'} a_{\alpha'}:\}|0\rangle,
\end{eqnarray}
\begin{eqnarray}
N_{11}(\alpha:\alpha')&=&\langle 0|\{a^+_\alpha, a_{\alpha'}\}|0\rangle=\delta_{\alpha\alpha'}, \\
N_{12}(\alpha:\alpha'\beta'\gamma)&=&\langle 0|\{a^+_\alpha, :a^+_{\gamma'} a_{\beta'} a_{\alpha'}:\}|0\rangle=0,\\
N_{21}(\alpha\beta\gamma:\alpha')&=&\langle 0|\{:a^+_\alpha a^+_\beta a_\gamma:, a_{\alpha'}\}|0\rangle=0,
\\
N_{22}(\alpha\beta\gamma:\alpha'\beta'\gamma'&)&
\nonumber \\
=\langle 0|\{:a^+_\alpha a^+_\beta a_\gamma:&,&:a^+_{\gamma'} a_{\beta'} a_{\alpha'}:\}|0\rangle.
\label{norm22}
\end{eqnarray}
Here, $\{~\}$ implies the anticommutator, $\{A,B\}=AB+BA$.
The norm matrix $N_{22}$ is given in Appendix B.
The matrix elements in Eq.~(\ref{hrpa1})
can be expressed by using those in Eq.~(\ref{hrpa}) such as
\begin{eqnarray}
A=a\times N_{11}\\
B=b\times N_{11},\\
C=c\times N_{22}.
\end{eqnarray}
The matrix $D$ consists of the terms of two types, one expressed by 
$D_1=d\times N_{22}$ and the other given by $D_2$, which originates from the terms with five fermion operator $:a^+_{\lambda_5} a^+_{\lambda_4} a^+_{\lambda_3} a_{\lambda_2} a_{\lambda_1}:$ 
in $[H,:a^+_\alpha a^+_\beta a_\gamma:]$:
\begin{eqnarray}
 &[&H,:a^+_\alpha a^+_\beta a_\gamma:]=\sum_{\alpha'}b(\alpha\beta\gamma:\alpha')a^+_{\alpha'}
\nonumber \\
 &+&\sum_{\alpha'\beta'\gamma'}d(\alpha\beta\gamma:\alpha'\beta'\gamma'):a^+_{\alpha'}a^+_{\beta'}a_{\gamma'}:
\nonumber \\ 
 &+&\sum_{\lambda_1\lambda_2\lambda_3\lambda_4\lambda_5}e(\alpha\beta\gamma:\lambda_1\lambda_2\lambda_3\lambda_4\lambda_5)
 \nonumber \\
&\times&
:a^+_{\lambda_1}a^+_{\lambda_2}a^+_{\lambda_3}a_{\lambda_5}a_{\lambda_4}:.
\label{d2}
\end{eqnarray}
Using an additional norm matrix
\begin{eqnarray}
N_{32}&(&\lambda_1\lambda_2\lambda_3\lambda_4\lambda_5:\alpha\beta\gamma)
\nonumber \\
&=&
\langle 0|\{:a^+_{\lambda_1} a^+_{\lambda_2} a^+_{\lambda_3} a_{\lambda_5} a_{\lambda_4}:, :a^+_\gamma a_\beta a_\alpha:\}|0\rangle,
\end{eqnarray}
$D_2$ can be expressed as $e\times N_{32}$. Thus $D_2$ expresses the coupling to five fermion configurations and 
includes the self-energy contributions to the three-fermion configurations in $Y_{\alpha\beta:\gamma}^\mu$ \cite{ts13}.
In the numerical applications $|0\rangle$ is approximated by the TDDM ground state and
the three-body correlation matrix $C_3$ contained in $N_{32}$ is expressed  by the squares of $C_2$ \cite{toh20}.

The transition amplitudes are given by
\begin{eqnarray}
\left(
\begin{array}{c}
x^{\mu}\\
X^{\mu}
\end{array}
\right)
=
\left(
\begin{array}{cc}
N_{11}&N_{12}\\
N_{21}&N_{22}
\end{array}
\right)
\left(
\begin{array}{c}
y^{\mu}\\
Y^{\mu}
\end{array}
\right).
\end{eqnarray}

\subsubsection{Symmetry properties}
Here it is shown that the Hamiltonian matrix of Eq.~(\ref{hrpa1}) is hermitian.
For this purpose the operator identity
\begin{eqnarray}
\langle 0|\{[H,\hat{A}],\hat{B}\}|0\rangle&+&\langle 0|\{[H,\hat{B}],\hat{A}\}|0\rangle
\nonumber \\
&=&\langle 0|[H,\{\hat{A},\hat{B}\}]|0\rangle
\label{jacobi}
\end{eqnarray}
is used.
In the matrix $A$ of Eq.~(\ref{hrpa1}), the operators $\hat{A}$ and $\hat B$ 
are $a^+_\alpha$ and $a_{\alpha'}$, respectively. Since
$\{\hat{A},\hat{B}\}$ is unity, the right-hand side
of Eq.~(\ref{jacobi}) vanishes, which means $\langle 0|\{[H,\hat{A}],\hat{B}\}|0\rangle=-\langle 0|\{[H,\hat{B}],\hat{A}\}|0\rangle$
and 
\begin{eqnarray}
A(\alpha:\alpha')^* &=&-\langle 0|\{[H,a_\alpha], a^+_{\alpha'}\}|0\rangle,
\nonumber \\
&=&\langle 0|\{[H,a^+_{\alpha'}],a_\alpha\}|0\rangle
=A(\alpha':\alpha).
\end{eqnarray}
In the case of the matrix $B$ in Eq.~(\ref{hrpa1}) $\hat{A}$ is $:a^+_\alpha a^+_\beta a_\gamma:$ and $\hat{B}$ is $a_{\alpha'}$, and
$\{\hat{A},\hat{B}\}$ is reduced to a one-body operator. Due to the ground-state condition in TDDM
\begin{eqnarray}
i\dot{n}_{\alpha\alpha'}&=&\langle 0|[a^+_{\alpha'}a_\alpha,H]| 0\rangle=0,
\label{tddm1}
\end{eqnarray}
the right-hand side
of Eq.~(\ref{jacobi}) vanishes, which means $\langle 0|\{[H,\hat{A}],\hat{B}\}|0\rangle=-\langle 0|\{[H,\hat{B}],\hat{A}\}|0\rangle$
and 
\begin{eqnarray}
B(\alpha\beta\gamma:\alpha')^* &=&-\langle 0|\{[H,:a^+_\gamma a_\beta a_\alpha :], a^+_{\alpha'}\}|0\rangle,
\nonumber \\
&=&\langle 0|\{[H,a^+_{\alpha'}], :a^+_\gamma a_\beta a_\alpha:\}|0\rangle
\nonumber \\
&=&C(\alpha':\alpha\beta\gamma).
\end{eqnarray}
Similarly, for the matrix $D$ in Eq.~(\ref{hrpa1}) $\hat{A}$ is $:a^+_\alpha a^+_\beta a_\gamma:$ and $\hat{B}$ is
$:a^+_{\gamma'} a_{\beta'} a_{\alpha'}$, and $\{\hat{A},\hat{B}\}$ is reduced to at most a two-body operator. Due to the
ground-state conditions Eq. (\ref{tddm1}) and
\begin{eqnarray}
i\dot{C}_{\alpha\beta\alpha'\beta'}&=&\langle 0|[:a^+_{\alpha'}a^+_{\beta'}a_\beta a_\alpha:,H]|0\rangle=0,
\label{tddm2}
\end{eqnarray}
the right-hand side of Eq.~(\ref{jacobi}) vanishes, which implies
\begin{eqnarray}
D(\alpha\beta\gamma:\alpha'\beta'\gamma')^*=D(\alpha'\beta'\gamma':\alpha\beta\gamma).
\end{eqnarray}
Therefore, the Hamiltonian matrix in Eq. (\ref{hrpa1}) is hermitian. This means that the left-hand eigenvectors $((\bar{y}^\mu)^*~~(\bar{Y}^\mu)^*)$ are
equivalent to $(({y}^\mu)^*~~({Y}^\mu)^*)$ and that the right and left transition amplitudes satisfy similar conjugate relations $(\chi^\mu~~{\cal X}^\mu)=((x^\mu)^*~~(X^\mu)^*)$, 
leading to 
\begin{eqnarray}
\sum_\mu X^\mu_{\alpha'\beta':\beta}\chi^\mu_{\alpha}&=&\sum_\mu (x^\mu_{\alpha}{\cal X}^\mu_{\alpha'\beta':\beta})^*.
\end{eqnarray}
When the truncation of higher amplitudes is made in the calculations of $X^\mu_{\alpha'\beta':\beta}$ and ${\cal X}^\mu_{\alpha\beta:\beta'}$, however,
the relation
\begin{eqnarray}
\sum_\mu (x^\mu_{\alpha}{\cal X}^\mu_{\alpha'\beta':\beta})^*=\sum_\mu x^\mu_{\alpha'}{\cal X}^\mu_{\alpha\beta:\beta'}
\end{eqnarray}
that implies symmetry under the exchange $(\alpha,\beta)\leftrightarrow(\alpha',\beta')$
does not hold exactly and, consequently, Eq. (\ref{cond0}) is not completely fulfilled.

It has been pointed out in Ref. \cite{ts13} that the left-hand eigenvectors $(({y}^\mu)^*~~({Y}^\mu)^*)$ corresponds to the amplitudes 
for a particle state with $N+1$.
Thus Eq.~(\ref{hrpa1}) gives simultaneously the particle states and the hole states. 
This is completely analogous to pp(hh)RPA (see Ref. \cite{RS}.)

\subsubsection{Self energy}
In the following it is shown that the self-energy of the Green's function is derived from the equation for the transition amplitude $x^\mu$.
By using $y^\mu=x^\mu$ and $Y^\mu=(N_{22})^{-1}X^\mu$ Eq. (\ref{hrpa1}) for $Y^\mu$ is expressed as 
\begin{eqnarray}
By^\mu+DY^\mu=bx^\mu+D(N_{22})^{-1}X^\mu=\omega_\mu X^\mu.
\end{eqnarray}
This gives
\begin{eqnarray}
X^\mu&=&\frac{1}{\omega_\mu-D(N_{22})^{-1}}bx^\mu
\nonumber \\
&=&\frac{1}{\omega_\mu-D(N_{22})^{-1}}N_{22}v x^\mu,
\end{eqnarray}
where $b=C^*=N_{22}v$ is used.
By inserting this into Eq. (\ref{stddm1}) the equation for $x^\mu$ is written as 
\begin{eqnarray}
(\epsilon_\alpha-\omega_\mu)x^\mu_\alpha+v\frac{1}{\omega_\mu-D(N_{22})^{-1}}N_{22}v x^\mu
=0,
\label{self}
\end{eqnarray}
where the sum over single-particle indices is implicit.
The matrix
\begin{eqnarray}
h(\omega)=\epsilon_\alpha +v\frac{1}{\omega-D(N_{22})^{-1}}N_{22}v,
\label{homega}
\end{eqnarray}
can be regarded as an effective Hamiltonian for $x^\mu$ and the second term gives the self-energy of the Green's function. $h(\omega)$ is hermitian in EoRPA. 
An equation similar to Eq. (\ref{self}) is used to obtain the eigenstates in the Green's function method based on the PVC approach \cite{litv}.

\subsubsection{Hartree-Fock approximation for the ground state}
If the HF approximation is taken for the ground state $|0\rangle$, 
$N_{22}$ in Eq. (\ref{norm22}) becomes
\begin{eqnarray}
N_{22}(\alpha\beta\gamma&:&\alpha'\beta'\gamma')=
(\delta_{\alpha\alpha'}\delta_{\beta\beta'}-\delta_{\alpha\beta'}\delta_{\beta\alpha'})\delta_{\gamma'\gamma}
\nonumber \\
&\times&(n^0_{\gamma\gamma}+n^0_{\alpha\alpha}n^0_{\beta\beta}-n^0_{\gamma\gamma}n^0_{\alpha\alpha}-n^0_{\gamma\gamma}n^0_{\beta\beta}),
\end{eqnarray}
where $n^0_{\alpha\alpha}$ is equal to 1 or 0. In HF, $N_{22}$ is non-vanishing only for $Y_{\rm pp':h}^\mu$ and $Y_{\rm hh':p}^\mu$. 
Here h and p refer to a hole state and a particle state, respectively.
These amplitudes $Y_{\rm pp':h}^\mu$ and $Y_{\rm hh':p}^\mu$ correspond to the backward amplitudes of $y_{\rm h}^\mu$ and $y_{\rm p}^\mu$, respectively.
Hereafter this formulation consisting of the four amplitudes, 
$y_{\rm h}^\mu$, $y_{\rm p}^\mu$, $Y_{\rm pp':h}^\mu$ and 
$Y_{\rm hh':p}^\mu$ is referred to 
as oRPA. 
For completeness the full expression of oRPA is given
\begin{eqnarray}
(\epsilon_\alpha&-&\omega_\mu)y^\mu_\alpha
\nonumber \\
&+&\sum_{\rm pp'h}\langle {\rm pp'}|v|\alpha {\rm h}\rangle{\tilde Y}^\mu_{\rm pp':h}
\nonumber \\
&+&\sum_{\rm hh'p}\langle {\rm hh'}|v|\alpha {\rm p}\rangle{\tilde Y}^\mu_{\rm hh':p}=0,
\label{orpa1}
\end{eqnarray}
\begin{eqnarray}
(\epsilon_{\rm p}+\epsilon_{\rm p'}-\epsilon_{\rm h}&-&\omega_\mu){\tilde Y}^\mu_{\rm pp':h}
\nonumber \\
&+&\sum_{\alpha}\langle \alpha {\rm h}|v|{\rm pp'}\rangle_A y^\mu_\alpha
\nonumber \\
&+&\sum_{\rm p_1p_2}\langle {\rm p_1p_2}|v|{\rm pp'}\rangle{\tilde Y}^\mu_{\rm p_1p_2:h}
\nonumber \\
&+&\sum_{\rm p_1h_1}[\langle {\rm hp_1}|v|{\rm ph_1}\rangle_A{\tilde Y}^\mu_{\rm p_1p':h_1}
\nonumber \\
&+&\langle {\rm hp_1}|v|{\rm p'h_1}\rangle_A{\tilde Y}^\mu_{\rm pp_1:h_1}]=0,
\label{orpa21}
\\
(\epsilon_{\rm h}+\epsilon_{\rm h'}-\epsilon_{\rm p}&-&\omega_\mu){\tilde Y}^\mu_{\rm hh':p}
\nonumber \\
&+&\sum_{\alpha}\langle \alpha {\rm p}|v|{\rm hh'}\rangle_A y^\mu_\alpha
\nonumber \\
&-&\sum_{\rm h_1h_2}\langle {\rm h_1h_2}|v|{\rm hh'}\rangle{\tilde Y}^\mu_{\rm h_1h_2:p}
\nonumber \\
&-&\sum_{\rm p_1h_1}[\langle {\rm ph_1}|v|{\rm hp_1}\rangle_A{\tilde Y}^\mu_{\rm h_1h':p_1}
\nonumber \\
&+&\langle {\rm ph_1}|v|{\rm h'p_1}\rangle_A{\tilde Y}^\mu_{\rm hh_1:p_1}]=0,
\label{orpa22}
\end{eqnarray}
where $y^\mu_\alpha$ denotes $y^\mu_{\rm h}$ or $y^\mu_{\rm p}$, and ${\tilde Y}^\mu_{\alpha\beta:\gamma}$
means $Y^\mu_{\alpha\beta:\gamma}-Y^\mu_{\beta\alpha:\gamma}$. The Hamiltonian matrix for Eqs. (\ref{orpa1})--(\ref{orpa22}) is hermitian
because Eqs. (\ref{tddm1}) and (\ref{tddm2}) are satisfied by the HF condition and by $C_{\alpha\beta\alpha'\beta'}=0$.
In this approximation ${Y}^\mu_{\rm pp':h}$ and ${Y}^\mu_{\rm hh':p}$ do not couple to other three-fermion amplitudes. Therefore, oRPA corresponds to the Green's function approach,
TDA of Ref. \cite{hrpa96}.
\begin{figure} 
\begin{center} 
\includegraphics[height=6cm]{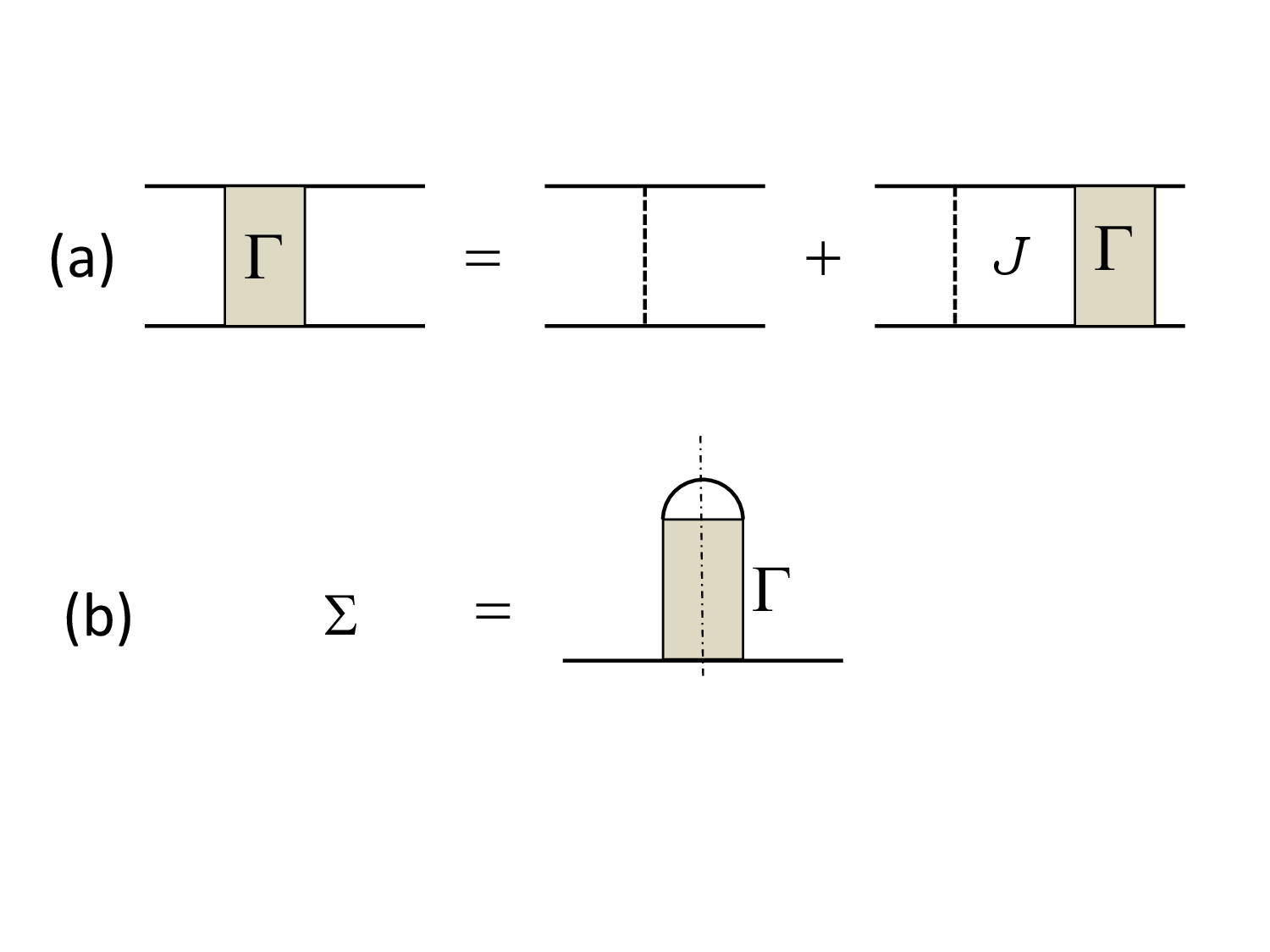}
\end{center}
\caption{(a) Equation for the vertex function $\Gamma$ (the squares with $\Gamma$). The horizontal lines depict the Green's functions, the dash line means the residual interaction and $J$ the product of the two
Green's functions. (b) The self-energy $\Sigma$. The thick lines depict the Green's functions. The vertical dot-dashed line in (b) which intersects three-fermion lines represents $Y_{\rm pp':h}^\mu$ or $Y_{\rm hh':p}^\mu$.} 
\label{ggt} 
\end{figure}
If correlations in ${Y}^\mu_{\rm pp':h}$ and ${Y}^\mu_{\rm hh':p}$ are restricted to p--p and h--h pairs, oRPA corresponds to the Green's function approaches, the $G_0G_0t-$matrix \cite{pieri} and pp-RPA \cite{sogo,urban}. 
In these Green's function approaches the self-energy $\Sigma$ is approximated by $G_0\Gamma$,
where $G_0$ is the HF single-particle Green's function and $\Gamma$ the vertex function. The equation for $\Gamma$ and the expression for $\Sigma$ 
are schematically shown in Fig. \ref{ggt} (a) and (b), respectively. In Fig. \ref{ggt} 
the thick lines depict the Green's functions, the dash line means the residual interaction and $J$ the product of two
Green's functions. 
In $G_0G_0t-$matrix \cite{pieri} and pp-RPA \cite{urban} all the Green's functions are given by $G_0$. 
The vertical dot-dashed line in Fig. \ref{ggt} (b) which intersects three-fermion lines corresponds to $Y_{\rm pp':h}^\mu$ or $Y_{\rm hh':p}^\mu$. Since $\Sigma$ is given by $G_0\Gamma$,
the correlations between the two fermions, one in $G_0$ and the other in $\Gamma$ are not considered in $G_0G_0t-$matrix and pp-RPA. 
In oRPA these correlations are included as the p--h correlations in $Y_{\rm pp':h}^\mu$ and $Y_{\rm hh':p}^\mu$.
In the PVC models the p--h correlations included in $Y_{\rm pp':h}^\mu$ and $Y_{\rm hh':p}^\mu$ are expressed by the RPA phonons \cite{colo,litv} but p--p and h--h correlations are not included.

\begin{figure} 
\begin{center} 
\includegraphics[height=6cm]{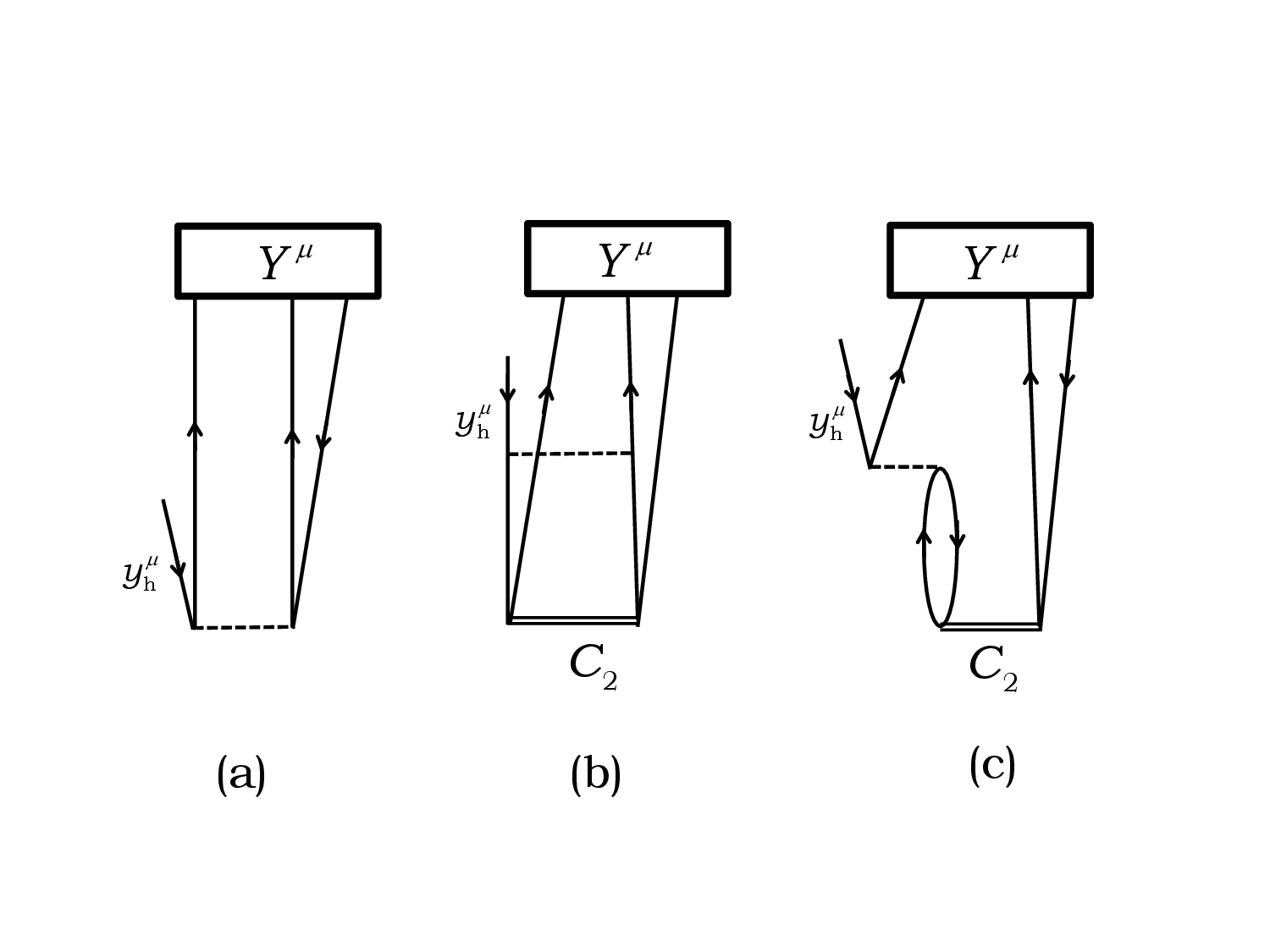}
\end{center}
\caption{Coupling of $y^\mu_{\rm h}$ to $Y^\mu_{\rm pp':h'}$. The vertical lines with arrows denote either a hole state or a particle state, the dashed lines mean the residual interaction,
the squares with $Y^\mu$ show $Y^\mu_{\rm pp':h'}$, and the horizontal bars with $C_2$ the correlation matrix. The process (a) is included in oRPA, and the processes (b) and (c) are additionally included in EoRPA.} 
\label{yY} 
\end{figure}

As will be shown in Sect.III, the condition Eq. (\ref{cond0})  
\begin{eqnarray}
\sum_{\mu}y_{\rm p}^\mu({Y}_{\rm hh':p'}^\mu)^*=\sum_{\mu}(y_{\rm h}^\mu)^* Y_{\rm pp':h'}^\mu
\label{HF}
\end{eqnarray}
is not satisfied in oRPA except for special cases such as the paring Hamiltonian \cite{rich} where the hole space is symmetric to the particle space.
This may be understood in the following way. The two-body correlation matrix $C_{\rm pp'hh'}$, for example, includes all p--p, h--h and p--h correlations.
On the other hand $\sum_{\mu}y_{\rm p}^\mu({Y}_{\rm hh':p'}^\mu)^*$ includes full h--h and partial p--h correlations but no p--p correlations are included as seen from Eq. (\ref{orpa22})
for ${Y}_{\rm hh':p'}^\mu$.
Similarly, $({y}_{\rm h}^\mu)^* Y_{\rm pp':h'}^\mu$ includes full p--p and partial p--h correlations but no h--h correlations are included (see Eq. (\ref{orpa21})). This asymmetry causes the violation of Eq. (\ref{HF}).
In EoRPA $C_2$ included in Eq. (\ref{hrpa1}) plays a role in reducing the asymmetry between particle-space and hole-space correlations.
The coupling of $y^\mu_{\rm h}$ to $Y^\mu_{\rm pp':h'}$ included in $B$ (Eq. (\ref{hrpa1})) is shown in Fig. \ref{yY}. 
The vertical lines with arrows denote either a hole state or a particle state, the dashed lines mean the residual interaction,
the squares with $Y^\mu$ show $Y^\mu_{\rm pp':h'}$, and the horizontal bars with $C_2$ the two-body correlation matrix. The process (a) is included both in oRPA and in EoRPA.
The processes (b) and (c) are additionally included in EoRPA. These processes are also called vertex corrections. As mentioned above, $C_2$ in (b) and (c) includes full p--p, h--h and p--h correlations
and thus can supplement h--h correlations missing in the process (a).
The two-body correlation matrix $C_2$ included in $D$ and the three-body correlation matrix $C_3$ in $D_2$ also represent additional correlations that are not included in oRPA.

When a correlated ground state is used, other three-fermion amplitudes such as $Y_{\rm hh':h''}^\mu$ and $Y_{\rm ph:p'}^\mu$ ($Y_{\rm pp':p''}^\mu$ and $Y_{\rm ph:h'}^\mu$) which are the backward amplitudes of
$Y_{\rm pp':h}^\mu$ ($Y_{\rm hh':p}^\mu$)
have non-vanishing values of $N_{22}$, and such a formulation with the backward amplitudes may correspond to FRPA of Ref. \cite{dick02}.
However, the inclusion of such three-fermion amplitudes with small values of $N_{22}$ sometimes causes unphysical fragmentation of the single-particle strength \cite{hrpa96,ts13}.
Therefore, the same four amplitudes $y_{\rm h}^\mu$, $y_{\rm p}^\mu$, $Y_{\rm pp':h}^\mu$ and 
$Y_{\rm hh':p}^\mu$ as those in oRPA are used also in EoRPA. 

\subsubsection{Renormalization approximation}
Here the approximation where the ground-state correlation effects are partially included through $n_{\alpha\alpha'}$ neglecting $C_2$ is considered.
Such a treatment of ground-state correlations may be called the renormalization approximation \cite{Rowe}. This approximation in the Green's function method is called r-RPA \cite{durel}.
When $C_2$ is neglected, $N_{22}$ in Eq. (\ref{norm22}) becomes
\begin{eqnarray}
N_{22}(\alpha\beta\gamma:\alpha'\beta'\gamma')&=&
(\delta_{\alpha\alpha'}\delta_{\beta\beta'}-\delta_{\alpha\beta'}\delta_{\beta\alpha'})\delta_{\gamma'\gamma}
\nonumber \\
&\times&N(\alpha\beta\gamma)
\end{eqnarray}
where $N(\alpha\beta\gamma)=n_{\gamma}+n_{\alpha}n_{\beta}-n_{\gamma}n_{\alpha}-n_{\gamma}n_{\beta}$. Here, $n_{\alpha\alpha'}=\delta_{\alpha\alpha'}n_\alpha$ is assumed.
The Hamiltonian matrix is given by
\begin{eqnarray}
A(\alpha:\alpha')&=&\epsilon_\alpha\delta_{\alpha\alpha'},\\
B(\alpha\beta\gamma:\alpha')&=&C(\alpha':\alpha\beta\gamma)^*
\nonumber \\
&=&\langle\alpha'\gamma|v|\alpha\beta\rangle_AN(\alpha\beta\gamma),
\label{r-RPA0}
\\
D(\alpha\beta\gamma:\alpha'\beta'\gamma')&=&\delta_{\gamma'\gamma}[(\epsilon_\alpha+\epsilon_\beta-\epsilon_\gamma)
\nonumber \\
&\times&(\delta_{\alpha\alpha'}\delta_{\beta\beta'}-\delta_{\alpha\beta'}\delta_{\beta\alpha'})
\nonumber \\
&+&\langle \alpha'\beta'|v|\alpha\beta\rangle_A(1-n_\alpha-n_\beta)]
\nonumber \\
&\times&N(\alpha'\beta'\gamma'),
\label{r-RPA}
\end{eqnarray}
where the p--h correlations in $D$ are omitted similarly to r-RPA.
$D$ is not hermitian because $(1-n_\alpha-n_\beta)$ differs from $N(\alpha\beta\gamma)$. Removing the term $n_\alpha n_\beta$ from $N(\alpha\beta\gamma)$ can make the Hamiltonian matrix hermitian.
This approximation corresponds to r-RPA \cite{durel}.
The condition Eq. (\ref{cond0}) is not fulfilled in the renormalization approximation because of the same reason as in oRPA and EoRPA, and the number conservation law is violated \cite{durel}.

\section{Applications}
\subsection{Pairing model}
First the 
pairing Hamiltonian \cite{rich} is considered to show that EoRPA is a reasonably good approximation. The pairing Hamiltonian is given as
\begin{eqnarray}
H=\sum_{i=1}^\Omega\epsilon_\alpha(a^+_i a_i+a^+_{\bar{i}}a_{\bar{i}})
-g\sum_{i\neq j}^\Omega a^+_i a^+_{\bar{i}}a_{\bar{j}}a_j.
\label{hpair}
\end{eqnarray}
Here $g$ is the strength of the pairing force acting in a space of $\Omega (=N)$ twofold degenerate equidistant orbitals
with the single-particle energies $\epsilon_i=(i-1)\Delta\epsilon$. This Hamiltonian has extensively been used to 
investigate the validity of theoretical approaches \cite{s21}. 

In Table \ref{tab0} the occupation probabilities calculated in TDDM for $N=4$ and $g/\Delta \epsilon=0.5$ are compared with the results in exact diagonalization approach (EDA).
In HF the $i=1$ and $2$ states are fully occupied and the $i=3$ and $4$ states are empty.
The TDDM results agree well with the EDA results. 
\begin{table}
\caption{Single-particle energies $\epsilon_i$ and occupation probabilities 
$n_i$ calculated in TDDM for the pairing model with $N=4$ and $g/\Delta \epsilon=0.5$. The results in EDA are given in parentheses.}
\begin{center}
\begin{tabular}{ccc} \hline
$i$ &{$\epsilon_i/\Delta \epsilon$}&{$n_{i}$}\\ \hline
$1$ & 0 & 0.964 (0.965) \\
$2$ & 1 & 0.909 (0.901)  \\
$3$ & 2 & 0.091 (0.099)  \\
$4$ & 3 & 0.036 (0.035) \\\hline
\end{tabular}
\label{tab0}
\end{center}
\end{table}

\begin{figure} 
\begin{center} 
\includegraphics[height=6cm]{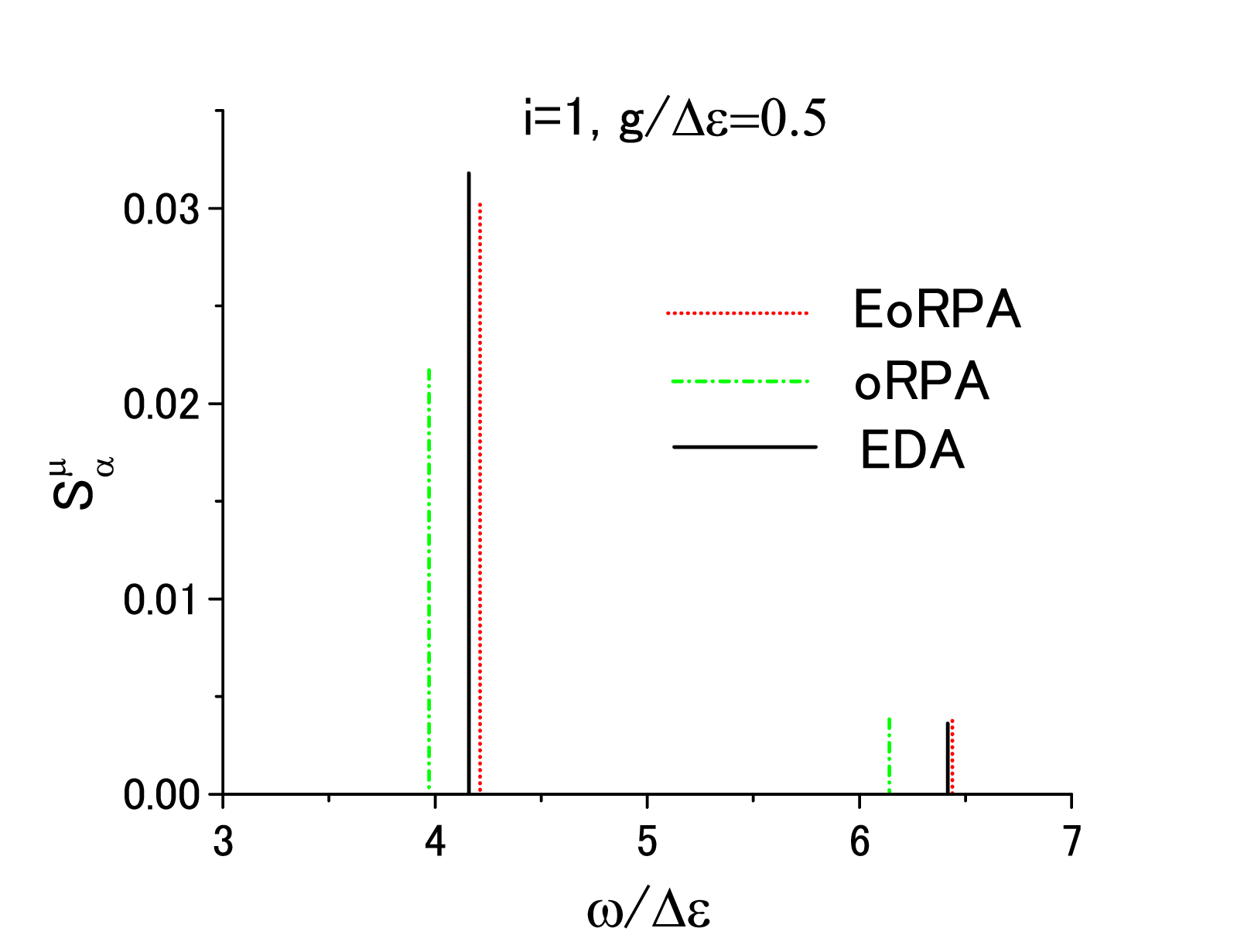}
\end{center}
\caption{Transition strength of the $i=1$ state in $N=4\rightarrow 5$ 
calculated in EoRPA with the TDDM ground state for the pairing model with $g/\Delta \epsilon=0.5$ (the dotted lines). The results in oRPA with the HF ground state 
are depicted with the dot-dashed lines.
The EDA results are shown with the solid lines.} 
\label{pick1} 
\end{figure}

\begin{figure} 
\begin{center} 
\includegraphics[height=6cm]{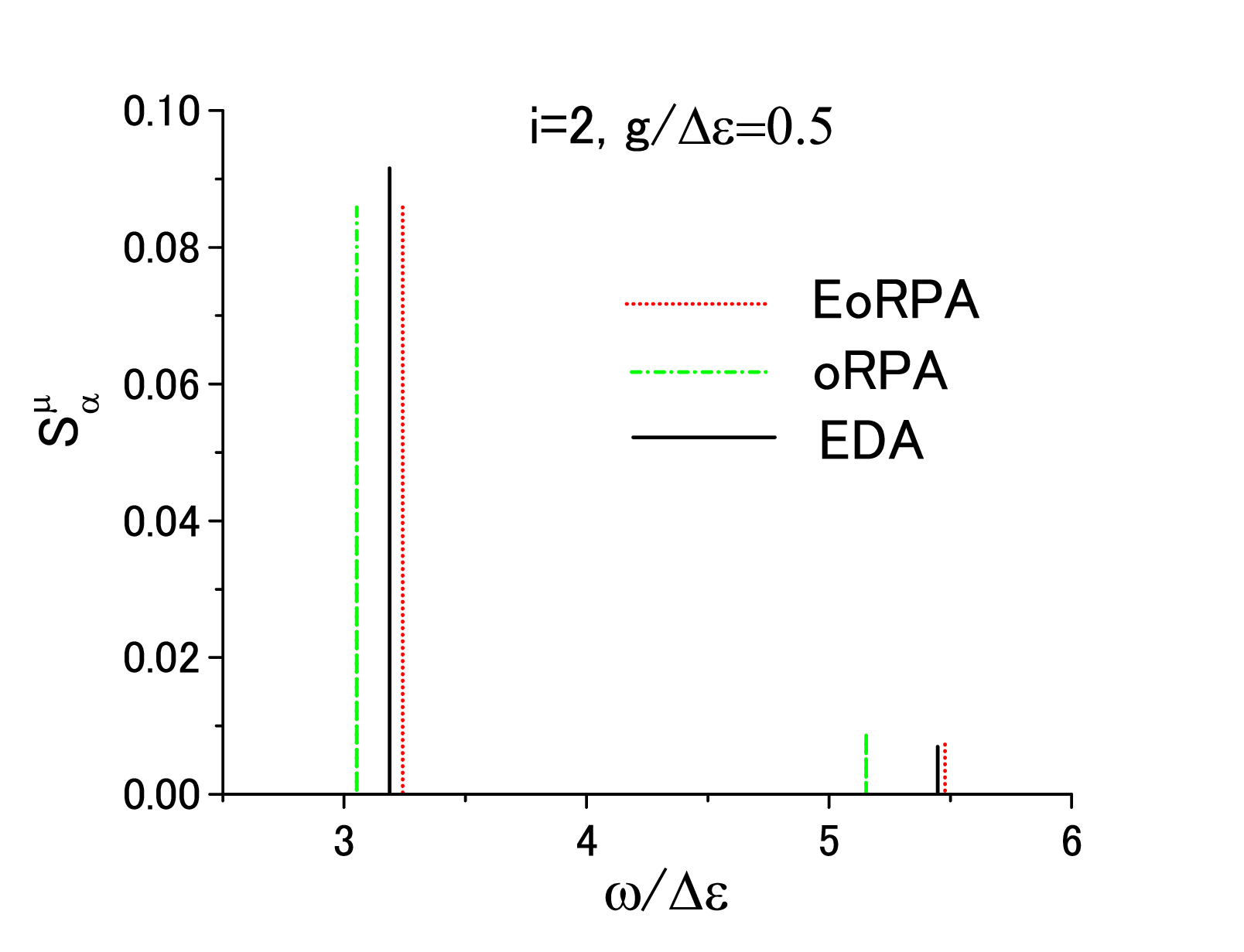}
\end{center}
\caption{Same as Fig. \ref{pick1} but for the $i=2$ state.} 
\label{pick2} 
\end{figure}

The transition strength $S_\alpha^\mu= x^\mu_{\alpha}(x^\mu_\alpha)^*$ of the $i=1$ state in $N=4\rightarrow 5$ calculated in EoRPA (dotted lines) is shown in Fig. \ref{pick1}. 
Since the $i=1$ state in the $N=4$ system is not fully occupied due to the ground-state correlations, such a transition is allowed,
and Eq. (\ref{hrpa1}) can handle both the single-particle removal and addition processes.
Since the interaction in the pairing Hamiltonian Eq. (\ref{hpair}) is allowed for p--p or h--h pairs, $y^\mu_{\rm h}$ can couple only to $Y^\mu_{\rm pp':h}$.
The results in oRPA and EDA are shown with the dot-dashed and solid lines, respectively.
In EDA the single-particle transition strength is calculated by using the ground state for the $N=4$ system and the eigenstates for the $N=5$ system.
The peaks around $\omega/\Delta \epsilon =4$ are due to the coupling to $Y^\mu_{3\bar{3}:1}$ whose unperturbed energy is $4~\Delta \epsilon$ and those above 
$\omega/\Delta \epsilon =6$ are from the coupling to $Y^\mu_{4\bar{4}:1}$ whose unperturbed energy is $6~\Delta \epsilon$. 
The two peaks in EoRPA are shifted upward as compared with the oRPA results and become closer to the results in EDA.
This difference between the EoRPA and oRPA is mainly attributed to the self-energy contributions included in the $D$ matrix in EoRPA.
The self-energy contributions play a role in effectively increasing the energies of the three-fermion configurations in $Y^\mu_{\rm pp':h}$.
The transition strength of the $i=2$ state in $N=4\rightarrow 5$ calculated in oRPA (dot-dashed lines) and EoRPA (dotted lines) is shown in Fig. \ref{pick2}. 
The peaks near $\omega/\Delta \epsilon =3$ are due to the coupling to $Y^\mu_{3\bar{3}:2}$ whose unperturbed energy is $3~\Delta \epsilon$ and those above 
$\omega/\Delta \epsilon =5$ are from the coupling to $Y^\mu_{4\bar{4}:2}$ whose unperturbed energy is $5~\Delta \epsilon$. 
As in the case of the $i=1$ state the two peaks in EoRPA are shifted upward as compared with the oRPA results and become closer to the results in EDA (solid lines).
The transition strength distributions of the $i=3$ and $4$ states in $N=4\rightarrow 3$ are the same as those of the $i=2$ and $1$ states, respectively, and are not shown here.
The occupation probability of each single-particle state is shown in Table \ref{tabp}.
The occupation probabilities of the $i=1$ and $2$ are obtained from $1-\sum_\mu S^\mu_\alpha$, where $S^\mu_\alpha$ is the transition strength shown in Figs. \ref{pick1} and \ref{pick2}.
The results in EoRPA agree well with those in TDDM.
Due to the hole--particle symmetry of the pairing Hamiltonian, the occupation probabilities of the $i=3$ and $4$ states are equal to $1-n_2$ and $1-n_1$, respectively,
and the the number conservation law is fulfilled both in oRPA and EoRPA as shown in the bottom line of Table \ref{tabp}, where $N=\sum_\alpha n_\alpha$.
\begin{table}
\caption{Occupation probabilities 
$n_{\alpha}$ calculated in TDDM, EoRPA and oRPA for the pairing Hamiltonian at $g/\Delta \epsilon=0.5$.}
\begin{center}
\begin{tabular}{ccccc} \hline
 i &{$\epsilon_i/\Delta \epsilon$}&TDDM&EoRPA&oRPA\\ \hline
$1$ & 0 & 0.964 &0.96606&0.97444 \\
$2$ & 1 & 0.909 &0.90685&0.91706  \\
$3$ & 2 & 0.091 &0.09315& 0.08294\\
$4$ & 3 & 0.036 &0.03394&0.02556 \\\hline
&$N$ & 4&4&4\\\hline
\end{tabular}
\label{tabp}
\end{center}
\end{table}

\subsection{$^{16}$O}
In this section EoRPA is applied to the single-particle states in $^{16}$O where hole-particle symmetry is broken. 
It is shown that the number conservation law is drastically improved by going from oRPA to EoRPA.
The small single-particle space consisting of the proton $1p_{3/2}$, $1p_{1/2}$ and $1d_{5/2}$ states is used
to facilitate the comparison with the results in EDA. In this single-particle space 
$y^\mu_{\rm h}$ and $y^\mu_{\rm p}$ respectively couple only to the backward amplitudes $Y^\mu_{\rm pp':h}$ ad $Y^\mu_{\rm hh':p}$ due to parity selection rule.
These couplings to the backward amplitudes are essential to determine the occupation probabilities below and above the Fermi level \cite{ts13}.
The single-particle energies and wavefunctions are obtained from the HF calculation with the Skyrme III force (SIII).
Only the $t_0$ term of SIII is used as the residual interaction. The value of $t_0$ is reduced by 40 \% 
to obtain ground-state correlations comparable to the results of other theoretical calculations \cite{adachi,utsuno}.
The TDDM ground state of $^{16}$O is calculated by using the adiabatic method where $C_3$ is approximated by the square of $C_2$ \cite{toh20}. 
The occupation probabilities are shown in Table \ref{tab1} where the results
in EDA are given in parentheses. In HF the $1p_{3/2}$ and $1p_{1/2}$ states are full occupied and the $1d_{5/2}$ state is empty.
TDDM well reproduce the results in EDA.
\begin{table}
\caption{Proton single-particle energies $\epsilon_\alpha$ and occupation probabilities 
$n_{\alpha}$ calculated in TDDM for $^{16}$O. The results in EDA are given in parentheses.}
\begin{center}
\begin{tabular}{ccc} \hline
 Orbit&{$\epsilon_\alpha$ [MeV]}&{$n_{\alpha}$}\\ \hline
$1p_{3/2}$ & -18.2 & 0.924 (0.922) \\
$1p_{1/2}$ & -12.0 & 0.836 (0.822)  \\
$1d_{5/2}$ & ~-3.8 & 0.106 (0.111)  \\\hline
\end{tabular}
\label{tab1}
\end{center}
\end{table}

\begin{figure} 
\begin{center} 
\includegraphics[height=6cm]{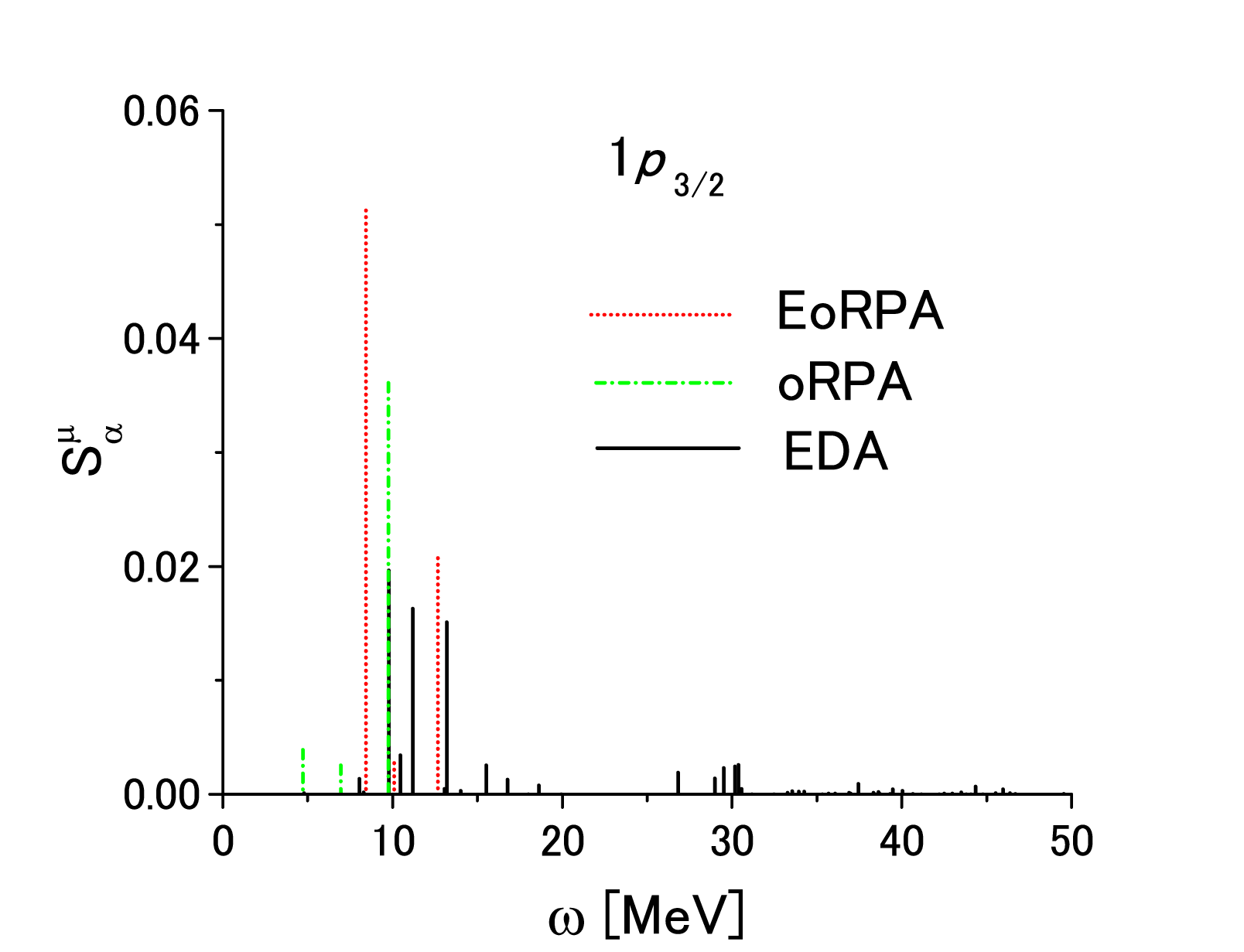}
\end{center}
\caption{Transition strength of the proton $1p_{3/2}$ state in $^{16}$O$\rightarrow$$^{17}$F 
calculated in EoRPA with the TDDM ground state for $^{16}$O (dotted lines). The results in oRPA with the HF ground state for $^{16}$O 
are depicted with the dot-dashed lines.
The EDA results are shown with the solid lines.} 
\label{p3} 
\end{figure}

\begin{figure} 
\begin{center} 
\includegraphics[height=6cm]{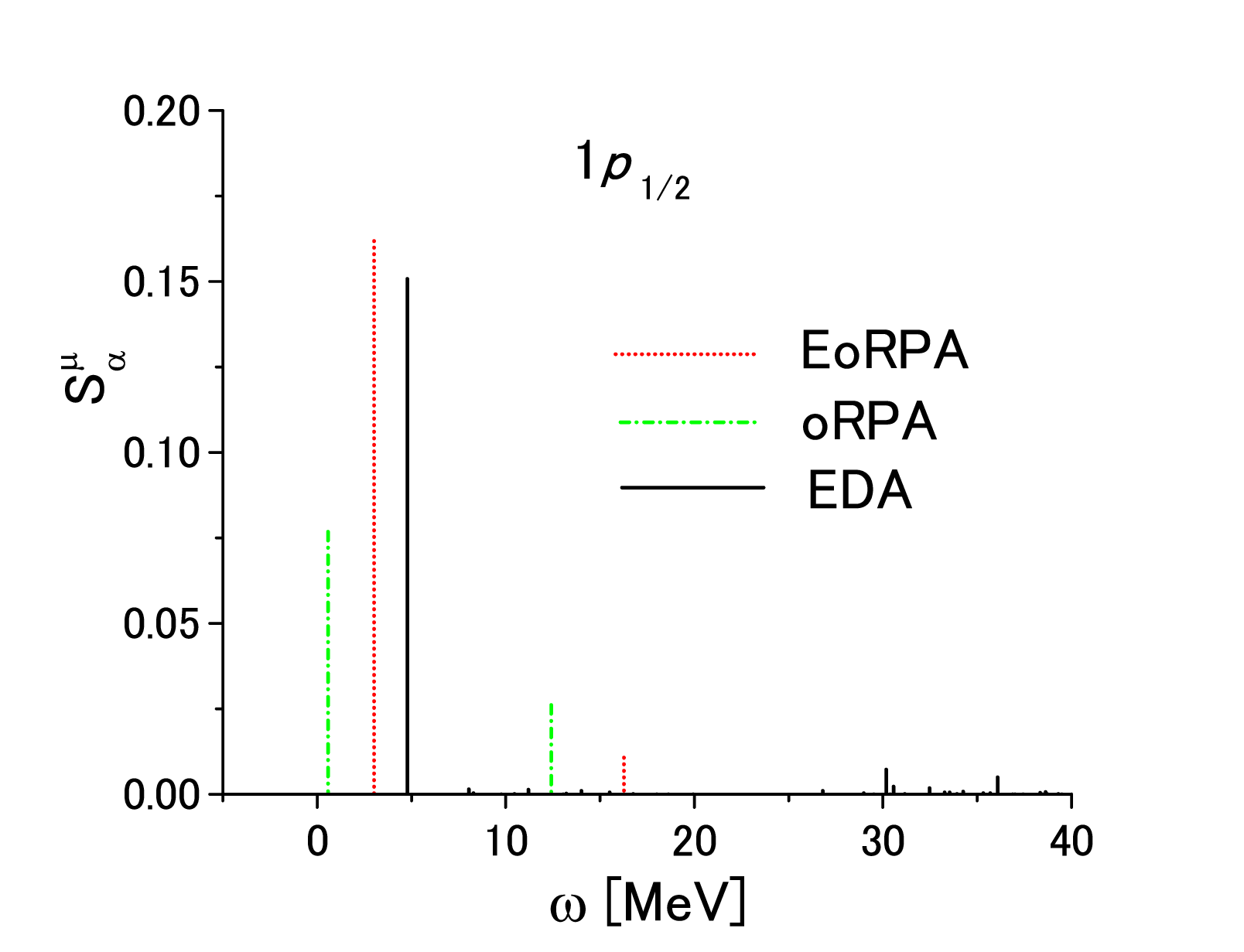}
\end{center}
\caption{Same as Fig. \ref{p3} but for the proton $1p_{1/2}$ state.} 
\label{p1} 
\end{figure}

The transition strength of the proton $1p_{3/2}$ state in $^{16}$O$\rightarrow$$^{17}$F calculated in EoRPA (dotted lines) is shown in Fig. \ref{p3}.
Since the proton $1p_{3/2}$ state in $^{16}$O is not fully occupied due to the ground-state correlations, such a single proton transfer is allowed, and
Eq. (\ref{hrpa1}) can describe both the single-particle removal and addition processes.
The results in oRPA and EDA are shown with the dot-dashed and solid lines, respectively.
In EDA the single-particle transition strength is calculated by using the ground state for the $N=6$ system and the eigenstates for the $N=7$ system.
The small peaks seen below 10 MeV in the oRPA result disappear in EoRPA and the EoRPA strength is fragmented around 10 MeV, which is consistent with the EDA result.
This difference between the EoRPA and oRPA may be explained by the self-energy contributions included in the $D$ matrix in EoRPA.
However, EoRPA and also oRPA cannot produce the strength distribution in high-energy region seen in the EDA results. 
Apparently higher amplitudes such as $Y^\mu_{\rm p_1p_2p_3p_4: h_1h_2h_3}$ 
whose unperturbed energies are from 20 to 40 MeV are needed to explain the strength in the high-energy region.
The five fermion amplitudes $Y^\mu_{\rm h_1h_2h_3: p_1p_2}$ can couple to $Y^\mu_{\rm pp': h}$ but would have negligible effects because their unperturbed energies are smaller than $-28$ MeV.
The transition strength of the proton $1p_{1/2}$ state in $^{16}$O$\rightarrow$$^{17}$F calculated in EoRPA (dotted lines) is shown in Fig. \ref{p1}.
As in the case of the $1p_{3/2}$ state, the peaks in EoRPA are shifted upward from the oRPA positions and locate closer to the EDA results but again
EoRPA cannot produce the small strength distribution in high-energy region seen in the EDA results.

\begin{figure} 
\begin{center} 
\includegraphics[height=6cm]{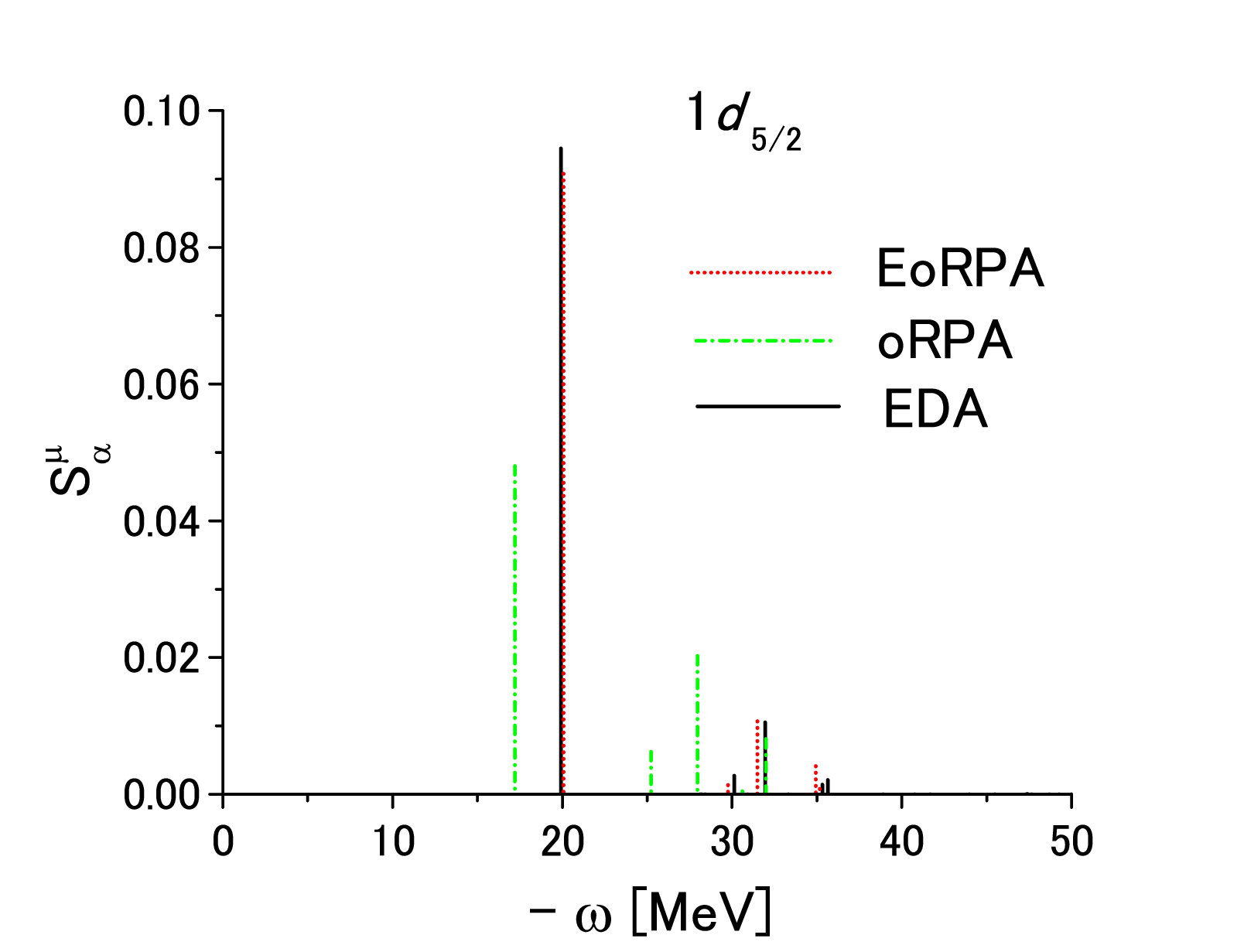}
\end{center}
\caption{Same as Fig. \ref{p3} but for the proton $1d_{5/2}$ state in $^{16}$O$\rightarrow$$^{15}$N.} 
\label{d5} 
\end{figure}

The transition strength of the proton $1d_{5/2}$ state in $^{16}$O$\rightarrow$$^{15}$N calculated in EoRPA (dotted lines) is shown in Fig. \ref{d5}.
Since such transition strength appears in the negative energy region \cite{ts13} because of $E_0<E_\mu$, it is shown as a function of $-\omega$.
In the case of a particle removal process the excitation energy is usually expressed as a negative quantity. 
The partial occupation of the proton $1d_{5/2}$ state in $^{16}$O due to the ground-state correlations enables such a transition.
The single-particle transition strength in EDA is calculated by using the ground state for the $N=6$ system and the eigenstates for the $N=5$ system.
As in the case of the $1p_{3/2}$ state, the peaks in EoRPA are shifted upward and each main peak in EoRPA is very closer to that in EDA. 

\begin{table}
\caption{Occupation probabilities 
$n_{\alpha}$ calculated in TDDM, EoRPA and oRPA for $^{16}$O.}
\begin{center}
\begin{tabular}{ccccc} \hline
 Orbit&{$\epsilon_\alpha$ [MeV]}&TDDM&EoRPA&oRPA\\ \hline
$1p_{3/2}$ & -18.2 & 0.92361 &0.92524&0.95746 \\
$1p_{1/2}$ & -12.0 & 0.83619 &0.82733&0.89707  \\
$1d_{5/2}$ & ~-3.8 & 0.10553 &0.10771&0.08298 \\\hline
&$N$ & 6&6.00188&6.12186\\\hline
\end{tabular}
\label{tab2}
\end{center}
\end{table}

The number conservation law is discussed below. The occupation probabilities of the single-particle states $1p_{3/2}$ and $1p_{1/2}$
are given by $1-\sum_\mu S^\mu_\alpha$ where $S^\mu_\alpha$ is the strength shown in Figs. \ref{p3} and \ref{p1}. On the other hand the occupation probability of
the $1d_{5/2}$ state is equal to the sum of the strength in Fig. \ref{d5}. The occupation probabilities thus obtained are tabulated in Table \ref{tab2}, where
$N$ in the bottom line means 
$N=4\times n_{1p_{3/2}}+2\times n_{1p_{1/2}}+6\times n_{1d_{5/2}}$. The occupation probabilities in EoRPA agree well with those in TDDM.
In EDA the occupation probabilities obtained from the transition strengths shown in Figs. \ref{p3}--\ref{d5} are the same as those given in Table \ref{tab1}
and the total number is conserved.
The extent of the violation of the number conservation law is measured by $\Delta N/N=(N-6)/6$.
In oRPA $\Delta N/N$ is $2.03$ \%. 
This is of the same order of accuracy as TDA applied to $^{16}$O \cite{hrpa96} which gives $\Delta N/N=6.56$ \%. 
In EoRPA $\Delta N/N$ is drastically decreased to $0.031$ \%. 
Equation (\ref{hrpa}) that partially includes the effects of ground-state correlations (Eq. (\ref{hrpa}) 
is equivalent to Eq. (\ref{hrpa1}) without $D_2$) is also applied and $\Delta N/N$ is found $0.88$ \%.
The comparison between the oRPA and EoRPA results demonstrates that
the ground-state correlations included in EoRPA greatly alleviate the violation of the number conservation law.
In the pp-RPA like approximation that neglects the p--h correlations in oRPA $\Delta N/N$ is $3.29$ \%.
The pp-RPA Green's function formalism has been applied to boson-fermion mixtures \cite{sogo} and strongly polarized fermion gases \cite{urban} and it was
shown that the Luttinger theorem is fulfilled.
The only difference between oRPA without p--h correlations and the pp-RPA formalism is that
the latter includes backward going p--p and h--h correlations. This corresponds to the inclusion of the $Y^\mu_{\rm p_1p_2:p_3}$ and $Y^\mu_{\rm h_1h_2:h_3}$
amplitudes in oRPA. However, these amplitudes have zero values of $N_{22}$ in HF and cannot be included in oRPA.
In the r-RPA like treatment of EoRPA that uses the fractional occupation probabilities in Eqs. (\ref{r-RPA0}) and (\ref{r-RPA}) $\Delta N/N$ is decreased to $1.43$ \%
because the occupation factors $N(\alpha\beta\gamma)$ and $(1-n_\alpha-n_\beta)$ reduce p--p and h--h correlations. The r-RPA Green's function formalism \cite{durel} has also been applied to strongly polarized fermion gases
and it was found that the number conservation law is violated to the same extent as in the r-RPA like treatment of EoRPA.
It was shown in Ref. \cite{pieri} that the fully self-consistent Green's function method called the $GG t$-matrix or the Luttinger-Ward method where all $G_0$ including that closing the loop
in the self-energy $\Sigma$ (see Fig. \ref{ggt} (b)) are replaced by fully self-consistent $G$, the Luttinger theorem is exactly fulfilled with the full Dyson equation.
This might be due to the fact that such a self-consistent approach includes diagrams which can only be described by the higher amplitudes in EoRPA.
According to Ref. \cite{mahaux} only the approximation that includes the set of all the diagrams in the same order of the interaction strength is
number-conserving. The $GG t$-matrix does not satisfy the condition of Ref. \cite{mahaux} because 
the correlations between the two fermions, one outside and the other inside of the square box in Fig. \ref{ggt} (b) are not included in the self-consistent $GG t$-matrix.
Exact solutions include such correlations as a matter of course. The problem of the Luttinger theorem and approximate Green's function approaches seems to need further investigations.

The violation of number conservation is also related to the fact that the relation 
\begin{eqnarray}
C_{\alpha\beta\alpha'\beta'}&=&\sum_{\mu}({x}^\mu_{\alpha})^*{X}^{\mu }_{\alpha'\beta':\beta}
\label{c11}                              
\\
&=&\sum_{\mu}{x}^\mu_{\alpha'}({X}^{\mu }_{\alpha\beta:\beta'})^*
\label{c12}
\end{eqnarray}
does not exactly hold. For example, for $\alpha=\beta=1p_{1/2}$ and $\alpha'=\beta'=1d_{5/2}$ Eq. (\ref{c11}) gives $0.2200$ in EoRPA while Eq. (\ref{c12}) gives $0.2198$.
One may think that the violation of the number conservation law be caused by the inaccuracy of the TDDM ground state and the neglect of other three-fermion amplitudes in EoRPA.
So a calculation is made where $n_\alpha$, $C_2$ and $C_3$ calculated from the ground state in EDA are inserted into the EoRPA equation including all three-fermion amplitudes. The obtained value of
$\Delta N/N$ is 0.054\%, 
indicating that the accuracy of the ground state and the neglect of other three-fermion amplitudes in EoRPA is not the origin of the violation of the number conservation law.
In the single-particle space consisting of the $1p_{1/2}$, $1p_{3/2}$ and $1d_{5/2}$ states $y^\mu_{\rm h}$ does not couple to the forward amplitude 
$Y^\mu_{\rm hh':p}$ due to parity selection rule and also $y^\mu_{\rm p}$ has no coupling to $Y^\mu_{\rm pp':h}$. The couplings to the forward amplitudes $Y^\mu_{\rm hh':p}$ and $Y^\mu_{\rm pp':h}$ cause fragmentation 
of a hole state and a particle state, respectively.
To estimate the influence of such forward amplitudes on the number conservation law, additional oRPA and EoRPA calculations are performed where the proton $1s_{1/2}$ state is added to the single-particle space. 
The obtained occupation probabilities are shown in Table \ref{tab3}. The results in EoRPA agree well with those in TDDM and EDA, and $\Delta N/N$ is again drastically decreased by going 
from oRPA to EoRPA  but it is  of the same order of magnitude as the EoRPA result in Table \ref{tab2}. Thus the coupling to the forward amplitudes has little influence on the number conservation law.
\begin{table}
\caption{Occupation probabilities 
$n_{\alpha}$ calculated in TDDM, EoRPA and oRPA for $^{16}$O including the proton $1s_{1/2}$ state. The results in EDA are given in parentheses.}
\begin{center}
\begin{tabular}{ccccc} \hline
 Orbit&{$\epsilon_\alpha$ [MeV]}&TDDM&EoRPA&oRPA\\ \hline
$1s_{1/2}$ & -32.1 & 0.96924 (0.96873) &0.97068&0.99348 \\
$1p_{3/2}$ & -18.2 & 0.90814 (0.90553) &0.91105&0.95550 \\
$1p_{1/2}$ & -12.0 & 0.81183 (0.78752) &0.79981&0.89707  \\
$1d_{5/2}$ & ~-3.8 & 0.13422 (0.14423) &0.13653&0.10554 \\\hline
&$N$ & 8&8.00437&8.23635\\\hline
\end{tabular}
\label{tab3}
\end{center}
\end{table}
Presumably higher-amplitudes such as $X^\mu_{\rm p_1p_2p_3p_4: h_1h_2h_3}$ and $X^\mu_{\rm h_1h_2h_3h_4: p_1p_2p_3}$ which are not considered in EoRPA
are needed to further improve the number conservation law because ${x}^\mu_{\rm h_1}$ in $\sum_{\mu}({x}^\mu_{\rm h_1})^*{X}^{\mu  }_{\rm p_1p_2:h_2}$
and ${x}^\mu_{\rm p_1}$ in $\sum_{\mu}({x}^\mu_{\rm p_1})^*{X}^{\mu}_{\rm h_1h_2:p_2}$ (see Eq. (\ref{c11})) include the coupling to ${X}^{\mu  }_{\rm p_3p_4:h_3}$ and ${X}^{\mu  }_{\rm h_3h_4:p_3}$, respectively. 

\section{Summary}
The number conservation law in the odd-particle number random-phase approximation (oRPA) and its extension (EoRPA) was investigated
by applying them to the pairing model and $^{16}$O and comparing with the results in exact diagonalization approach (EDA).
It was found that the results in EoRPA generally have good agreement with those in EDA.
In the applications of oRPA and EoRPA to $^{16}$O it was shown that the number conservation law is not satisfied and that
it is drastically improved by going from oRPA to EoRPA due to the effects of ground-state correlations included in  EoRPA. 
It was pointed out that to fulfill the number conservation law the two-body correlation matrix consisting of the one-fermion and three-fermion transition amplitudes in EoRPA 
should be independent of how it is composed. Any extended oRPA approaches that are based on the truncation of higher transition amplitudes presumably cannot satisfy this condition, however.

\appendix
\section{Matrix elements of Eq. (\ref{hrpa})}
\begin{eqnarray}
a(\alpha:\alpha')=\epsilon_\alpha\delta_{\alpha\alpha'}
\end{eqnarray}
\begin{eqnarray}
b(\alpha\beta\gamma:\alpha')&=&\sum_{\lambda}\langle\alpha'\lambda|v|\alpha\beta\rangle_An_{\gamma\lambda}
\nonumber \\
&-&\sum_{\lambda\lambda'}[\langle\alpha'\lambda'|v|\alpha\lambda\rangle_A
(n_{\lambda\beta}n_{\gamma\lambda'}+C_{\gamma\lambda\lambda'\beta})
\nonumber \\
&+&\langle\alpha'\lambda'|v|\lambda\beta\rangle_A
(n_{\lambda\alpha}n_{\gamma\lambda'}+C_{\gamma\lambda\lambda'\alpha})
\nonumber \\
-\langle\alpha'\gamma|&v&|\lambda\lambda'\rangle_A(n_{\lambda\alpha}n_{\lambda'\beta}
+\frac{1}{2}C_{\lambda\lambda'\alpha\beta})],
\end{eqnarray}
\begin{eqnarray}
c(\alpha:\alpha'\beta'\gamma')&=&\langle\alpha'\beta'|v|\alpha\gamma'\rangle
\end{eqnarray}
\begin{eqnarray}
d&(&\alpha\beta\gamma:\alpha'\beta'\gamma')=(\epsilon_\alpha+\epsilon_\beta-\epsilon_\gamma)
\delta_{\alpha\alpha'}\delta_{\beta\beta'}\delta_{\gamma\gamma'}
\nonumber \\
&+&\frac{1}{2}\langle\alpha'\beta'|v|\alpha\beta\rangle_A\delta_{\gamma\gamma'}
\nonumber \\
&+&\sum_{\lambda}[\langle\lambda\alpha'|v|\alpha\gamma'\rangle_An_{\gamma\lambda}\delta_{\beta\beta'}
%\nonumber \\
-\langle\lambda\alpha'|v|\beta\gamma'\rangle_An_{\gamma\lambda}\delta_{\alpha\beta'}
\nonumber \\
&+&\langle\gamma\beta'|v|\lambda\gamma'\rangle_An_{\lambda\alpha}\delta_{\beta\alpha'}
%\nonumber \\
-\langle\gamma\beta'|v|\lambda\gamma'\rangle_An_{\lambda\beta}\delta_{\alpha\alpha'}
\nonumber \\
&-&\frac{1}{2}\delta_{\gamma\gamma'}(\langle\alpha'\beta'|v|\alpha\lambda\rangle_An_{\lambda\beta}
%\nonumber \\
+\langle\alpha'\beta'|v|\lambda\beta\rangle_An_{\lambda\alpha})].
\end{eqnarray}

\section{Norm matrix $N_{22}$}
The norm matrix $N_{22}$ is given as 
\begin{eqnarray}
N_{22}(\alpha\beta\gamma&:&\alpha'\beta'\gamma')=
(\delta_{\alpha\alpha'}\delta_{\beta\beta'}-\delta_{\alpha\beta'}\delta_{\beta\alpha'})n_{\gamma'\gamma}
\nonumber \\
&+&\delta_{\gamma\gamma'}(n_{\alpha\alpha'}n_{\beta\beta'}-n_{\alpha\beta'}n_{\beta\alpha'}
+C_{\alpha\beta\alpha'\beta'})
\nonumber \\
&-&\delta_{\alpha\alpha'}(n_{\gamma'\gamma}n_{\beta\beta'}+C_{\gamma'\beta\gamma\beta'})
\nonumber \\
&-&\delta_{\beta\beta'}(n_{\gamma'\gamma}n_{\alpha\alpha'}+C_{\gamma'\alpha\gamma\alpha'})
\nonumber \\
&+&\delta_{\alpha\beta'}(n_{\gamma'\gamma}n_{\beta\alpha'}+C_{\gamma'\beta\gamma\alpha'})
\nonumber \\
&+&\delta_{\beta\alpha'}(n_{\gamma'\gamma}n_{\alpha\beta'}+C_{\gamma'\alpha\gamma\beta'}).
\end{eqnarray}

\end{document}